\documentclass{article}
\usepackage{spconfa4,amsmath,graphicx}
\usepackage{cite}        
\usepackage{hyperref}    
\usepackage[nolist]{acronym}
\usepackage{amsfonts}
\usepackage[utf8]{inputenc}
\usepackage[T1]{fontenc}
\usepackage{tikz}
\usepackage{color,soul}
\usetikzlibrary{arrows.meta,positioning}
\usepackage{url}
\usepackage{multirow}
\title{PERSONALIZED CAUSAL LOW-BITRATE CODED SPEECH RESTORATION WITH SPEAKER EMBEDDINGS}
\name{%
  \parbox{\textwidth}{\centering\itshape
    Eray\,\"Ozg\"unay$^{1,2,3}$\thanks{Funded by the European Union under the Horizon Europe
    research and innovation programme, Grant Agreement No.~101119297 (EASYLI). 
    \newline
    \copyright~2026 IEEE. Author Accepted Manuscript. Licensed under \href{https://creativecommons.org/licenses/by/4.0/}{CC BY 4.0} with IEEE permission.},
    Christian\,Rollwage$^{1}$,
    Kishan\,Gupta$^{4}$,
    Andreas\,Brendel$^{4}$,\\[0.25em]
    Xavier\,Domont$^{3}$,
    Mattes\,Ohlenbusch$^{1}$,
    Jan\,Rennies$^{1,2}$,
    Simon\,Doclo$^{1,2}$%
  }%
}

\address{%
  \ninept
  \scalebox{0.98}[1]{%
    $^{1}$Fraunhofer Institute for Digital Media Technology IDMT, Oldenburg Branch for Hearing, Speech and Audio Technology HSA, Germany}%
  \\
  \ninept
  \scalebox{0.98}[1]{%
    $^{2}$Carl von Ossietzky Universit\"at Oldenburg, Dept. of Medical Physics and Acoustics and Cluster of Excellence Hearing4all, Germany}%
  \\
  \ninept
  $^{3}$CEOTRONICS AG, Germany\\
  \ninept
  $^{4}$Fraunhofer IIS, Germany%
}

\begin{document}
\ninept
\maketitle
\begin{acronym}
  \acro{CSR}{coded speech restoration}
  \acro{BWE}{bandwidth extension}
\end{acronym}

\begin{abstract}
Legacy low-bitrate speech codecs such as AMR-NB remain widely used in telephony, where they restrict speech bandwidth and degrade listening quality. Coded speech restoration (CSR) aims to restore the speech bandwidth while improving the overall quality of coded speech, but typically without adapting to different speakers. In this paper, we systematically investigate personalization for restoration of speech coded at low bitrates, using a causal, low-complexity Universal Bandwidth Extension Generative Adversarial Network (UBGAN) model as the basis. We propose two personalization methods based on speaker embeddings, which incur little additional computational cost at inference: (i) conditioning a single CSR model on speaker embeddings, and (ii) cluster-based model selection, where multiple personalized models are trained on clusters of acoustically similar speakers. Experiments are performed on AMR-NB coded speech at 4.75\,kbps. Experimental results with disjoint seen- and unseen-speaker test sets show that cluster-based model selection does not consistently outperform the non-personalized UBGAN baseline, whereas embedding conditioning significantly improves ViSQOL and WB-PESQ scores for both seen and unseen speakers.

\end{abstract}
\section{Introduction}
\label{sec:intro}

Despite recent advances in neural speech and audio coding\,\cite{kim24nsac}, low-resource telephony networks continue to rely on legacy low-bitrate codecs such as AMR-NB\,\cite{3gpp26071}, which restrict the coded speech bandwidth, degrade listening quality, and contribute to listener fatigue\,\cite{buthe25bwedemo}. Therefore, \ac{BWE} is typically included as part of \ac{CSR} \cite{schmidt23lpcgan, wen24ftcsr, gupta25ubgan}. However, \ac{BWE} can also magnify existing artifacts: distortions present in the lower band may propagate into the generated high-frequency content, potentially reducing overall quality \cite{gupta25ubgan}. For this reason, a target sampling rate of 16\,kHz is often an appropriate compromise between increasing speech bandwidth beyond narrowband telephony and avoiding excessive amplification of codec artifacts at higher target sampling rates. Nevertheless, even at 16\,kHz, residual codec artifacts and a loss of speaker-specific nuances may persist, raising the question of whether personalized \ac{CSR} can help mitigate these effects.

Personalization for \ac{CSR} is motivated by the success of personalization in related areas such as speech enhancement \cite{kolbaek16intelligibility, sivaraman20sparsemixture, sivaraman21zeroshot, eskimez22personalizedse, zmolikova23ntseoverview}, \ac{BWE} \cite{xu25boneconduction}, text-to-speech synthesis \cite{cooper20zeroshottts}, automatic speech recognition \cite{delcroix18speakerbeam, denisov19endtoend}, voice activity detection and diarization \cite{ding20personalvad, medennikov20tsvad}, and neural speech codecs \cite{jang24personalizedcodec}. In these works, two main methods are commonly explored. One line of research conditions a single model on speaker embeddings \cite{eskimez22personalizedse, xu25boneconduction, cooper20zeroshottts, zmolikova23ntseoverview, delcroix18speakerbeam, denisov19endtoend, ding20personalvad, medennikov20tsvad}, allowing a single model to adapt to different speakers. Following another research direction, multiple models are trained for small groups (clusters) of speakers, often combined with speaker-informed model selection at test time \cite{kolbaek16intelligibility, sivaraman20sparsemixture, sivaraman21zeroshot, jang24personalizedcodec}. Such clusters are typically obtained by grouping speakers in an embedding space. Because a personalized system can, in principle, exploit characteristic voice traits such as rhythm and timbre \cite{wang24usat}, it is worth investigating whether personalization can also improve \ac{CSR}. Despite this potential, to the best of our knowledge, personalization has not yet been systematically investigated for \ac{CSR}.

This paper investigates personalization for causal low-bitrate \ac{CSR} using the Universal Bandwidth Extension Generative Adversarial Network (UBGAN) \cite{gupta25ubgan} as a non-personalized \ac{CSR} basis, aiming to restore 8\,kHz AMR-NB coded speech at 4.75\,kbps to 16\,kHz wideband speech. At test time, we assume that a short clean enrollment utterance from the same speaker is available, from which a speaker embedding is computed. We consider and compare two alternative personalization methods for \ac{CSR}: (i) embedding conditioning, which conditions a single \ac{CSR} model on speaker embeddings; and (ii) cluster-based model selection, which trains a separate \ac{CSR} model per cluster. Both methods are designed to introduce only minimal additional computational cost to the CSR model at test time. Whereas \cite{ozgunay26lrac} explored these personalization methods with a non-causal HiFi-GAN+\,\cite{su21hifigan} basis, in this paper we investigate whether they also benefit a causal, low-delay, and low-complexity UBGAN CSR model, and additionally evaluate a hybrid combining both methods. An experimental comparison is provided on the VCTK dataset \cite{yamagishi2019vctk} for both seen speakers included in the training set and held-out unseen speakers.

\section{Method}
\label{sec:method}

\subsection{CSR Baseline}
\label{sec:ubgan_csr}
Let $u$ denote a clean reference utterance and $x$ the corresponding AMR-NB coded utterance. A \ac{CSR} system aims to reconstruct $u$ from $x$, yielding a restored 16\,kHz estimate $\hat{u}$.

The \ac{CSR} baseline is based on UBGAN\,\cite{gupta25ubgan}, which is a causal, low-complexity, low-delay model suitable for real-time processing on low-power devices. UBGAN operates in the subband domain using pseudo-quadrature mirror filter (PQMF) \cite{nguyen94pqmf} analysis/synthesis, a lightweight U-Net-like generator with temporal adaptive denormalization (TADE) \cite{mustafa21stylemelgan} residual blocks, and a multi-resolution STFT discriminator ensemble trained with a combination of adversarial, feature-matching, and reconstruction losses. In this paper, we follow the same subband-domain formulation: the coded utterance is first decomposed into PQMF subbands, a mel-spectrogram of the coded utterance $x$ is used as the only conditioning input to the generator (i.e., the blind UBGAN variant), and the processed subbands are resynthesized back to obtain the restored utterance. We also adapt UBGAN to restore 16\,kHz output from 8\,kHz AMR-NB coded speech, and allow the model not only to generate the missing high-frequency content but also to modify the existing lower-band subbands.

\subsection{Speaker Embeddings and Enrollment}
\label{sec:emb_and_enr}

All personalization methods considered here use speaker embeddings obtained from a pretrained ECAPA-TDNN\,\cite{desplanques20ecapatdnn} model for speaker verification, trained on the VoxCeleb1\,\cite{nagrani17voxceleb} and VoxCeleb2\,\cite{chung18b_voxceleb2} datasets and kept frozen in all experiments. The classification head is discarded and only the embedding part is used, yielding fixed-length 192-dimensional embedding vectors. Let $\mathcal{U}$ denote the set of all 16\,kHz clean reference utterances, and $\mathcal{X}$ the corresponding set of AMR-NB coded utterances. Let $f : \mathcal{U} \rightarrow \mathbb{R}^{192}$ denote the speaker embedder, which maps a clean utterance $u \in \mathcal{U}$ to a speaker embedding $\mathbf{e} = f(u)$.

\noindent\textbf{Training-time embeddings:}
For each speaker $s$ in the training set, let $\mathcal{U}_s^{\text{train}} \subset \mathcal{U}$ denote the set of clean utterances that are used as training targets. For each speaker $s$, a single training-time embedding is computed by averaging, i.e.
\begin{equation}
    \mathbf{e}_s^{\text{train}}
    = \frac{1}{\lvert \mathcal{U}_s^{\text{train}} \rvert}
      \sum_{u \in \mathcal{U}_s^{\text{train}}} f(u).
\end{equation}
The training-time embedding $\mathbf{e}_s^{\text{train}}$ is used both for embedding conditioning during training (Sec.\,\ref{sec:emb_cond}) and for defining the clusters in the cluster-based model selection method (Sec.\,\ref{sec:cluster}).

\noindent\textbf{Test-time embeddings:} For each speaker $s$ in the test set, let $\mathcal{U}_s^{\text{test}} \subset \mathcal{U}$ denote the set of clean utterances that are used as test targets, and let $\mathcal{X}_s^{\text{test}} \subset \mathcal{X}$ be the corresponding coded utterances. The $i$-th coded test utterance of speaker $s$ and its clean target are denoted by $x_{s,i}^{\text{test}} \in \mathcal{X}_s^{\text{test}}$ and $u_{s,i}^{\text{test}} \in \mathcal{U}_s^{\text{test}}$, respectively. For each target utterance $u_{s,i}^{\text{test}}$, we randomly select a different enrollment utterance $u_{s,j}^{\text{test}} \in \mathcal{U}_s^{\text{test}}$ from the same speaker with $j \neq i$; it should be noted that the same enrollment utterance $u_{s,j}^{\text{test}}$ may be used for multiple test utterances of the same speaker. The test-time embedding used for restoring $x_{s,i}^{\text{test}}$ is defined as $\mathbf{e}_{s,j}^{\text{test}} = f(u_{s,j}^{\text{test}})$. In contrast to the training-time embedding, which is a single per-speaker average, each test-time embedding is computed from one enrollment utterance. These test-time embeddings are used both for embedding conditioning at test time (Sec.\,\ref{sec:emb_cond}) and to select the cluster in the cluster-based model selection method (Sec.\,\ref{sec:cluster}).

In all cases, these embeddings are normalized to unit norm ($\lVert \mathbf{e} \rVert_2 = 1$).

\subsection{Embedding Conditioning}
\label{sec:emb_cond}

\begin{figure}[t]
    \centering
    \includegraphics[width=0.25\textwidth]{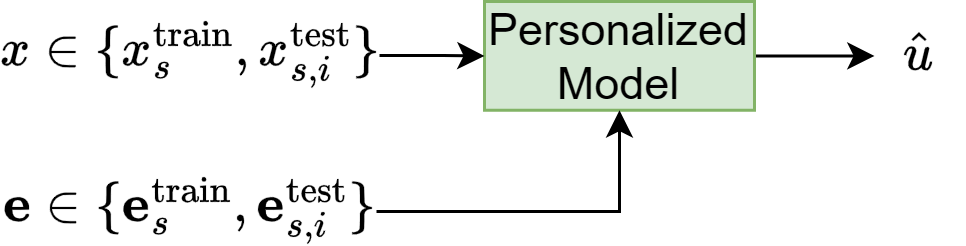}
    \caption{High-level view of embedding conditioning during training and test: the personalized model maps the coded utterance $x$ and the embedding $\mathbf{e}$ to the restored utterance $\hat{u}$.}
    \label{fig:conditioning_high_level}
\end{figure}

\begin{figure}[t]
    \centering
    \includegraphics[width=0.465\textwidth]{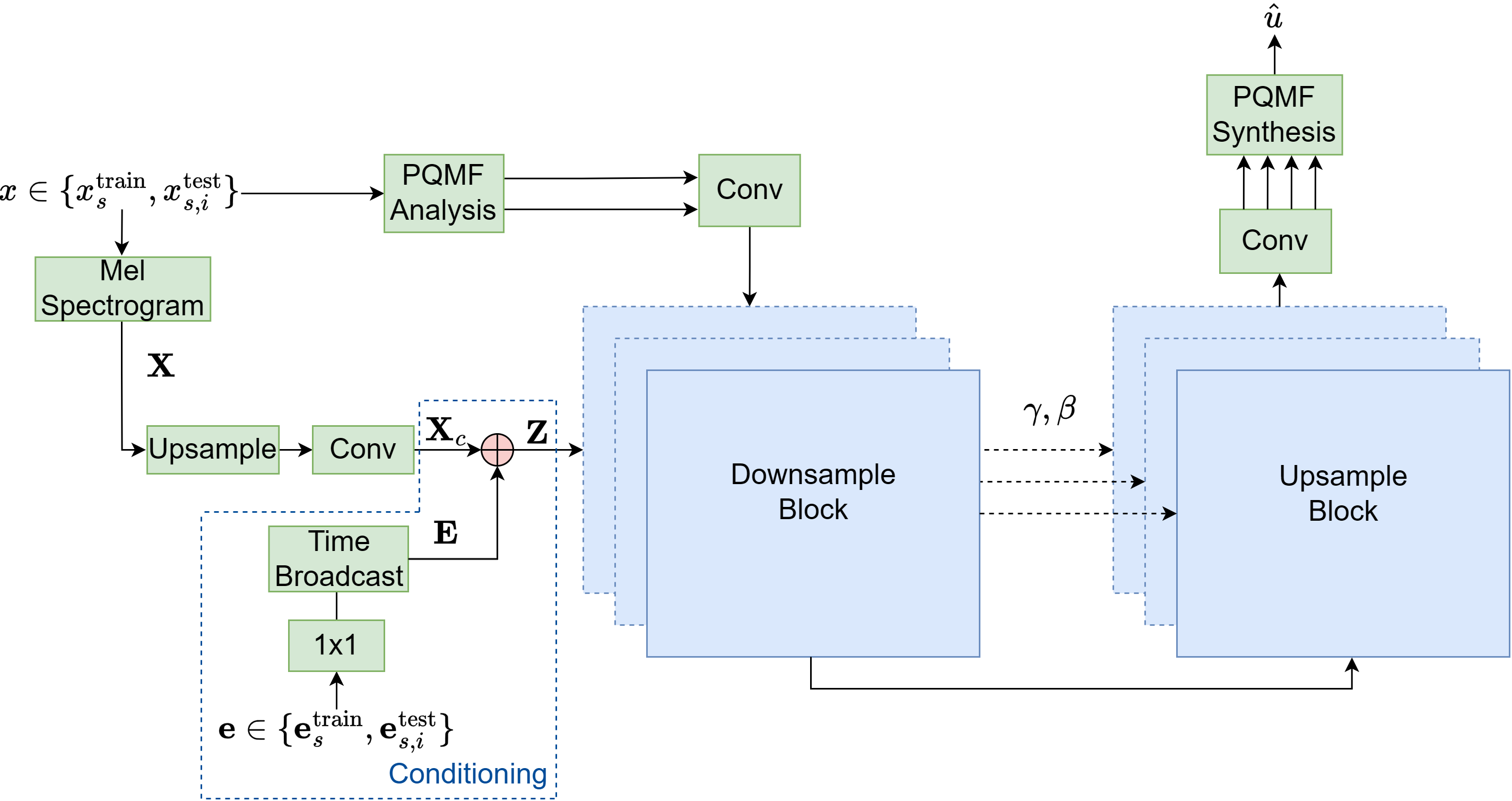}
    \caption{Low-level illustration of embedding conditioning in the UBGAN \ac{CSR} generator during training and test. The mel-spectrogram $\mathbf{X}$ of the coded input $x$ is processed by the conditioning pathway to obtain $\mathbf{X}_c$. The embedding $\mathbf{e}$ is mapped with a $1\times1$ convolution and repeated in time to obtain $\mathbf{E}$, which is added to $\mathbf{X}_c$. The sum $\mathbf{Z}$ is used as conditioning for the downsample blocks.}
    \label{fig:conditioning}
\end{figure}

The first proposed personalization method conditions a single \ac{CSR} model on a speaker embedding. High-level and low-level views of this method for UBGAN are shown in Fig.\,\ref{fig:conditioning_high_level} and Fig.\,\ref{fig:conditioning}, respectively. 

In the baseline, an 80-band mel-spectrogram $\mathbf{X}$ of the coded input utterance $x$ is computed and processed through upsampling and a convolutional layer, resulting in $\mathbf{X}_c$. This serves as the conditioning input for a stack of downsample blocks. In this method, these conditioning features are augmented with a speaker embedding. The 192-dimensional embedding $\mathbf{e}$ (see Sec.\,\ref{sec:emb_and_enr}) is adjusted to match the dimensions of $\mathbf{X}_c$ by mapping the channels with a learned $1\times1$ convolution and repeating along the temporal axis to obtain $\mathbf{E}$. Speaker conditioning is then performed by adding this time-broadcasted embedding to the mel-based conditioning features,
\begin{equation}
    \mathbf{Z} = \mathbf{X}_c + \mathbf{E},
\end{equation}
and the resulting $\mathbf{Z}$ is used as conditioning input to the downsample blocks. All subsequent processing inside the downsample and upsample blocks, as well as the PQMF analysis/synthesis and post-convolution, remain unchanged.

\subsection{Cluster-based Model Selection}
\label{sec:cluster}
The second proposed personalization method utilizes multiple \ac{CSR} models, each trained on a cluster of speakers that are close to each other in the speaker embedding space. The key idea is that speakers with similar embeddings are expected to share characteristic voice traits; training a separate model on each cluster allows it to focus on a narrower sub-task confined to a subset of speakers, similarly to cluster-specific neural speech codecs\,\cite{jang24personalizedcodec}.

\noindent\textbf{Training cluster-based models:}
All training-time (normalized) speaker embeddings $\mathbf{e}_s^{\text{train}}$ in Sec.\,\ref{sec:emb_and_enr} are partitioned into $C$ disjoint clusters using k-means in the embedding space; the resulting centroids are $\ell_2$-normalized. For each cluster, we train a separate \ac{CSR} model, architecturally identical to the baseline, using only the utterances from the speakers assigned to that cluster.

\begin{figure}[t]
    \centering
    \includegraphics[width=0.45\textwidth]{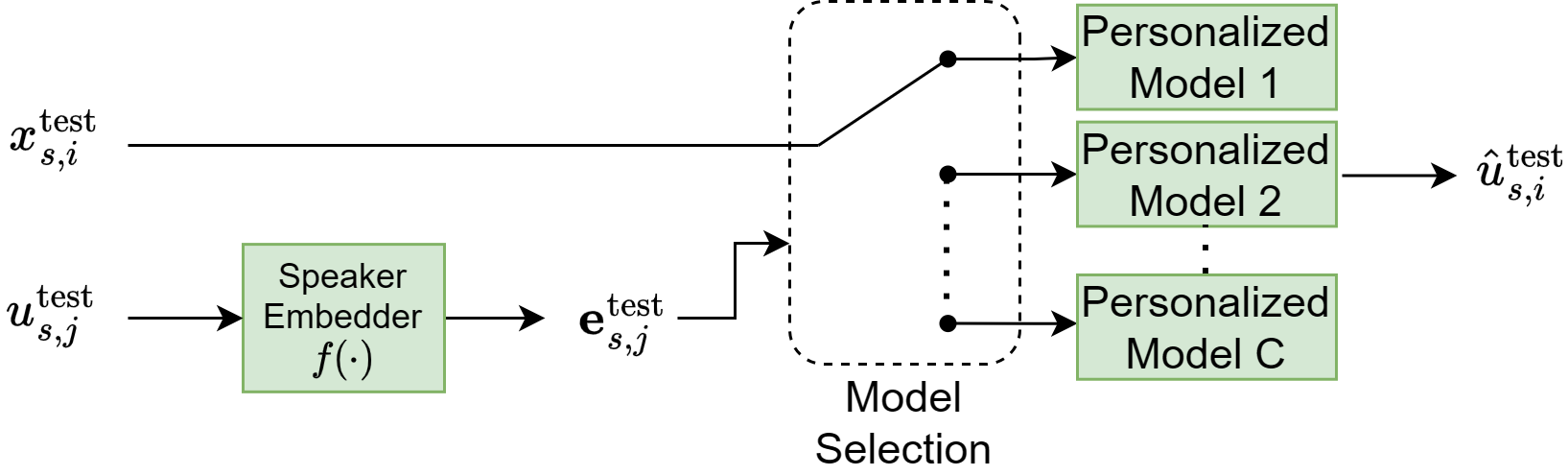}
    \caption{Cluster-based model selection at test time. For each coded test utterance $x_{s,i}^{\text{test}}$, a different clean enrollment utterance $u_{s,j}^{\text{test}}$ from the same speaker is passed through the speaker embedder to obtain $\mathbf{e}_{s,j}^{\text{test}} = f(u_{s,j}^{\text{test}})$. This embedding is assigned to the nearest cluster centroid, and the corresponding cluster-specific \ac{CSR} model is then used to generate the restored utterance $\hat{u}_{s,i}^{\text{test}}$ from $x_{s,i}^{\text{test}}$.}
    \label{fig:cluster_selection}
\end{figure}

\noindent\textbf{Test-time model selection:}
The cluster-based model selection at test time is illustrated in Fig.\,\ref{fig:cluster_selection}. For each coded test utterance $x_{s,i}^{\text{test}}$, the enrollment embedding $\mathbf{e}_{s,j}^{\text{test}}$ from the same speaker (computed on a different enrollment utterance $u_{s,j}^{\text{test}}$) is assigned to the closest cluster centroid using cosine distance, and the corresponding cluster-specific UBGAN model is selected.

Unlike the embedding conditioning method in Sec.\,\ref{sec:emb_cond}, speaker embeddings here are not used as conditioning inputs to the \ac{CSR} model; they serve only to define clusters during training and to select the appropriate cluster-specific model at test time.

As an additional experiment, we also evaluate a hybrid combining both methods, where each cluster-specific model is trained with embedding conditioning.

\section{Experimental Setup}
\label{sec:setup}

\subsection{Dataset and Splits}

Experiments are conducted on the VCTK dataset\,\cite{yamagishi2019vctk}, which comprises 110 English speakers with roughly 400 utterances per speaker (we exclude speaker p315 due to audio recording issues reported in\,\cite{yamagishi2019vctk}). All recordings are resampled to 16\,kHz and used as clean reference signals. The clean reference signals are downsampled to 8\,kHz and are coded with the AMR-NB codec at 4.75\,kbps using \texttt{ffmpeg}. The coded signals serve as input to the \ac{CSR} systems, while the original 16\,kHz signals are used as targets. Although our experiments target AMR-NB as a representative of legacy low-bitrate codecs, the proposed methods only require access to coded and clean references during training and are therefore, in principle, applicable to other codecs as well.

The speakers are split into three disjoint groups: 89 speakers for training, 10 speakers for validation, and 10 speakers held out entirely for testing on unseen speakers. For the 89 training speakers, we perform a per-speaker utterance split: for each training speaker, 80\% of their utterances are used for training and the remaining 20\% are held out for testing. The held-out utterances from these 89 speakers form the seen-speaker test set, while the utterances from the 10 held-out speakers form the unseen-speaker test set.

For the cluster-based model selection method (see Sec.\,\ref{sec:cluster}), we follow\,\cite{jang24personalizedcodec} and set the number of speaker clusters to $C = 4$. Applying k-means to the training-time embeddings of the 89 training speakers yields clusters of 16, 17, 22, and 34 speakers, respectively. Each of these clusters is used to train one personalized \ac{CSR} model.

\subsection{Training Configuration}
\label{sec:setup_training}

The architectural and training differences between the methods and the baseline are described in Secs.\,\ref{sec:emb_cond} and \ref{sec:cluster}. Despite the distribution mismatch between per-speaker averaged training-time embeddings and single-utterance test-time embeddings (see Sec.\,\ref{sec:emb_and_enr}), preliminary findings show that averaging outperforms single-utterance training-time embeddings, likely due to a more robust speaker representation.

Training is performed on randomly cropped 1\,s segments with a batch size of 64. The generator operates in the PQMF subband domain with a 2-band analysis frontend and a 4-band synthesis backend. Mel-spectrogram conditioning is computed directly from the coded input using 80 mel bands, an FFT size of 256, a Hann window length of 240 samples, a hop size of 160 samples, and a padding of 80 samples. The generator uses 6 down/up-sample blocks with factors $[1,2,2,2,2.5,2]$, 48 channels throughout, causal depthwise-separable convolutions with kernel size 7, and TADE-based residual modulation blocks. 

For adversarial training, we use a multi-resolution STFT discriminator ensemble with four sub-discriminators operating at FFT sizes 2048, 1024, 512, and 256, with corresponding hop sizes 512, 256, 128, and 64 and window lengths equal to the FFT sizes. The training objective combines a least-squares generative adversarial network (LSGAN) loss\,\cite{mao17lsgan}, a discriminator feature-matching loss, and reconstruction losses. The reconstruction term consists of a multi-resolution STFT loss with the same four STFT resolutions as above, together with an additional waveform-domain MSE loss. The feature-matching term is weighted by 10, while both the adversarial and reconstruction terms are weighted by 1.

We train the generator with the AdamW~\cite{loshchilov19adamw} optimizer and the discriminator with the Adam~\cite{kingma15adam} optimizer. The generator is trained with an initial learning rate of $5\times10^{-4}$, $(\beta_1,\beta_2) = (0.5,0.9)$, where $\beta_1$ and $\beta_2$ are the exponential decay rates for the first- and second-moment estimates of the optimizer, respectively, $\epsilon = 10^{-5}$, and weight decay 0.01. The discriminator is trained with a learning rate of $2\times10^{-4}$ with the same $\beta$ and $\epsilon$ values as for training the generator. The generator learning rate is decayed with an exponential scheduler with factor 0.99 once per epoch. Adversarial, feature-matching, and reconstruction losses are enabled from the start of training. In each training iteration, the discriminator is updated first and the generator is updated immediately afterwards. All models are trained for a maximum of 200{,}000 batch optimization steps. 

\subsection{Performance Metrics}

Testing is performed on both seen- and unseen-speaker sets by comparing the restored 16\,kHz output signals with their corresponding 16\,kHz clean reference signals. We consider three intrusive objective performance metrics: LSD\,\cite{lsd}, which quantifies spectral distortion (lower is better); WB-PESQ\,\cite{pesq}, computed using the Python \texttt{pesq} package (version\,0.0.4); and ViSQOL\,\cite{chinen20visqol} (version\,3.3.3) in speech mode at 16\,kHz. ViSQOL and WB-PESQ are perceptual similarity measures between reference and degraded speech (higher is better).

\subsection{Model Size and Complexity}
\label{sec:complexity}

\begin{table*}[t]
\centering
\caption{Model complexity and results for seen and unseen speakers (mean $\pm$ std).
Boldface indicates the best score for each metric within each baseline group.}
\resizebox{\textwidth}{!}{%
\begin{tabular}{llrc|ccc|ccc}
\hline
 & & & & \multicolumn{3}{c|}{Seen speakers} & \multicolumn{3}{c}{Unseen speakers} \\
Baseline & Method & Params & GMACs & LSD & ViSQOL & WB-PESQ & LSD & ViSQOL & WB-PESQ \\
\hline
\multirow{4}{*}{UBGAN}
 & —        & 309\,K             & 0.253 & 0.98 $\pm$ 0.04 & 2.80 $\pm$ 0.50 & 2.36 $\pm$ 0.35 & \textbf{0.97 $\pm$ 0.04} & 2.92 $\pm$ 0.52 & 2.28 $\pm$ 0.36 \\
 & Emb.\ cond.\       & 318\,K             & 0.253 & 0.98 $\pm$ 0.04 & \textbf{2.92 $\pm$ 0.50} & \textbf{2.47 $\pm$ 0.35} & 0.99 $\pm$ 0.05 & \textbf{2.97 $\pm$ 0.53} & \textbf{2.34 $\pm$ 0.35} \\
 & Cluster sel.\      & 4$\times$309\,K    & 0.253 & \textbf{0.97 $\pm$ 0.04} & 2.85 $\pm$ 0.50 & 2.33 $\pm$ 0.36 & 0.99 $\pm$ 0.05 & 2.86 $\pm$ 0.48 & 2.22 $\pm$ 0.34 \\
 & Hybrid             & 4$\times$318\,K    & 0.253 & 0.98 $\pm$ 0.05 & 2.88 $\pm$ 0.51 & 2.36 $\pm$ 0.35 & 1.00 $\pm$ 0.04 & 2.86 $\pm$ 0.50 & 2.24 $\pm$ 0.35 \\
\hline
\multirow{4}{*}{HiFi-GAN+}
 & —        & 1.06\,M            & 30.54 & 0.95 $\pm$ 0.03 & 2.76 $\pm$ 0.49 & 2.21 $\pm$ 0.32 & 0.94 $\pm$ 0.03 & 2.90 $\pm$ 0.53 & 2.18 $\pm$ 0.34 \\
 & Emb.\ cond.\       & 1.46\,M            & 30.54 & 0.95 $\pm$ 0.03 & 2.95 $\pm$ 0.51 & 2.22 $\pm$ 0.31 & 0.95 $\pm$ 0.03 & 3.01 $\pm$ 0.52 & 2.15 $\pm$ 0.32 \\
 & Cluster sel.\      & 4$\times$1.06\,M   & 30.54 & 0.92 $\pm$ 0.03 & \textbf{3.00 $\pm$ 0.48} & 2.39 $\pm$ 0.33 & \textbf{0.92 $\pm$ 0.03} & \textbf{3.06 $\pm$ 0.47} & 2.34 $\pm$ 0.33 \\
 & Hybrid             & 4$\times$1.46\,M   & 30.54 & \textbf{0.91 $\pm$ 0.03} & 2.99 $\pm$ 0.48 & \textbf{2.45 $\pm$ 0.32} & 0.93 $\pm$ 0.03 & 2.96 $\pm$ 0.48 & \textbf{2.37 $\pm$ 0.33} \\
\hline
\end{tabular}%
}
\label{tab:results}
\end{table*}

Table\,\ref{tab:results} includes the number of trainable generator parameters and multiply–accumulate operations (GMACs) per 1\,s segment of 8\,kHz narrowband input (8{,}000 samples). These figures exclude the ECAPA-TDNN speaker embedder, which in practice would need to be evaluated only once per speaker during an offline enrollment phase; the resulting embedding could then be stored and reused for all subsequent utterances, adding no computational cost to test-time \ac{CSR} inference. The embedding conditioning method introduces one additional $1\times1$ convolutional layer into the conditioning pathway, increasing complexity by only $0.028$\,MMACs (${\sim}0.01\%$ additional computation). For cluster-based model selection, each cluster-specific model shares the same architecture as the baseline; at test time only one model is evaluated, and cluster assignment requires computing the cosine distance between the enrollment embedding and $C$ centroids, which is negligible compared to the generator's inference cost. Both methods thus preserve the computational efficiency and low algorithmic delay of UBGAN\,\cite{gupta25ubgan}.

As an additional reference, Table\,\ref{tab:results} includes results from non-causal HiFi-GAN+-based \ac{CSR} systems reprinted from\,\cite{ozgunay26lrac} using identical data splits and evaluation pipeline, except for the hybrid method, which is newly evaluated here. HiFi-GAN+ is a WaveNet-based\,\cite{oord16wavenet} model with substantially higher complexity than the causal UBGAN baseline used here. For HiFi-GAN+, the embedding conditioning layer adds a larger relative overhead (1.46\,M vs.\ 1.06\,M parameters), and the hybrid requires storing $4\times1.46$\,M parameters in total.

\section{Results}
\label{sec:results}

Results for all methods are shown in Table\,\ref{tab:results}. For UBGAN, embedding conditioning consistently improves ViSQOL and WB-PESQ scores for both seen and unseen speakers, with LSD scores equal to the baseline for seen speakers but slightly worse for unseen speakers. Speaker-level paired Wilcoxon signed-rank tests confirm that the ViSQOL and WB-PESQ score improvements from embedding conditioning are significant for both seen ($p < 0.001$, $N = 89$) and unseen ($p < 0.02$, $N = 10$) speakers, suggesting that the model generalizes from the averaged training-time
embeddings rather than memorizing them, while the LSD score differences are not significant in either condition ($p > 0.05$). Cluster-based model selection, in contrast, shows mixed results: for seen speakers it slightly improves LSD and ViSQOL scores but reduces WB-PESQ scores, whereas for unseen speakers it is slightly worse than the baseline on all three metrics. One possible explanation is that UBGAN already handles speaker variability effectively with the full training set, leaving little room for cluster specialization to help, while the narrower speaker coverage per cluster may reduce generalization to unseen speakers. The hybrid method, which applies embedding conditioning within cluster-specific models, does not improve over embedding conditioning alone for UBGAN on any metric and, for unseen speakers, even falls below the non-personalized baseline on all three metrics.

Despite its substantially smaller size and lower complexity (see Sec.\,\ref{sec:complexity}), the UBGAN baseline already yields higher WB-PESQ and ViSQOL scores than the HiFi-GAN+\,\cite{su21hifigan} baseline from\,\cite{ozgunay26lrac}, albeit with slightly worse LSD scores. Notably, the best personalization method differs across baselines: with HiFi-GAN+, the hybrid and cluster-based model selection provide the largest gains, whereas with UBGAN, embedding conditioning is the most effective method. Since the two baselines differ in architecture, causality, and capacity, the relative benefit of each personalization method appears to depend on the underlying CSR model.

Across both baselines, the three metrics do not always agree, as LSD scores measure spectral distortion while ViSQOL and WB-PESQ scores assess perceptual quality. On average, ViSQOL scores are higher for unseen than for seen speakers for most systems, including the baselines. As ViSQOL is sensitive to the specific speech material\,\cite{chinen20visqol}, we hypothesize that this mainly reflects differences in the speakers and utterances across the two test sets rather than genuinely better performance on unseen speakers.

\section{Conclusion}
\label{sec:conclusion}

This paper investigated personalization for causal low-bitrate \ac{CSR} of AMR-NB coded speech using a UBGAN baseline with ECAPA-TDNN speaker embeddings, evaluated on the VCTK dataset. Cluster-based model selection yields mixed results and requires storing $C = 4$ models, and combining it with embedding conditioning in a hybrid approach does not improve over embedding conditioning alone for UBGAN. Embedding conditioning consistently improves ViSQOL and WB-PESQ scores for both seen and unseen speakers, maintaining low computational complexity. A comparison with a non-causal HiFi-GAN+ baseline further suggests that the most effective personalization method depends on the underlying CSR model. Notably, the causal UBGAN with embedding conditioning achieves WB-PESQ scores exceeding all HiFi-GAN+ configurations for seen speakers and ViSQOL scores surpassing the HiFi-GAN+ baseline, while requiring over $120\times$ fewer multiply–accumulate operations and enabling low-latency processing suitable for real-time deployment.

\bibliographystyle{IEEEbib}
\bibliography{refs}

@misc{ozgunay26lrac,
  author       = {Eray \"Ozg\"unay and Christian Rollwage and Jan Rennies and
                   Xavier Domont and Mattes Ohlenbusch and Simon Doclo},
  title        = {{Personalized Low-Bitrate Coded Speech Restoration via
                   Embedding Conditioning and Cluster-Based Model Selection}},
  howpublished = {presented at the {\it{Workshop on Low-Resource Audio Coding
                   (LRAC) at IEEE ICASSP}}, Barcelona, Spain},
  month        = may,
  year         = {2026},
  note         = {Available: \url{https://crowdsourcing.cisco.com/lrac-workshop/
        2026/assets/files/2026-LRAC-NAS-1-Paper-Ozgunay-Personalized-
        Low-Bitrate-Coded-Speech-Restoration-Via-Embedding-Conditioning-
        And-Cluster-Based-Model-Selection.pdf}}
}

@article{lsd,
  title={Distance measures for speech processing},
  author={Gray, Augustine and Markel, John},
  journal={IEEE Transactions on Acoustics, Speech, and Signal Processing},
  volume={24},
  number={5},
  pages={380--391},
  year={1976},
  publisher={IEEE}
}

@article{kim24nsac,
  author={Kim, Minje and Skoglund, Jan},
  journal={IEEE Signal Processing Magazine}, 
  title={{Neural Speech and Audio Coding: Modern AI technology meets traditional codecs}}, 
  year={2024},
  volume={41},
  number={6},
  pages={85-93},
}

@inproceedings{kingma15adam,
  author    = {Kingma, Diederik P. and Ba, Jimmy},
  title     = {{Adam: A Method for Stochastic Optimization}},
  booktitle = {Proc. 3rd Int. Conf. Learn. Representations (ICLR)},
  year      = {2015},
  month     = may,
  address   = {San Diego, CA, USA}
}

@inproceedings{loshchilov19adamw,
    title={Decoupled Weight Decay Regularization},
    author={Ilya Loshchilov and Frank Hutter},
    booktitle={Proc. International Conference on Learning Representations (ICLR)},
    year={2019},
    month= may,
    address = {New Orleans, LA, USA}
    }

@inproceedings{mao17lsgan,
  title={Least Squares Generative Adversarial Networks}, 
  author={Mao, Xudong and Li, Qing and Xie, Haoran and Lau, Raymond Y.K. and Wang, Zhen and Smolley, Stephen Paul},
  booktitle={Proc. IEEE International Conference on Computer Vision (ICCV)}, 
  year={2017},
  month = oct,
  address = {Venice, Italy}
}

@techreport{3gpp26071,
  author      = {{3rd Generation Partnership Project (3GPP)}},
  title       = {{Mandatory Speech Codec Speech Processing Functions; Adaptive Multi-Rate ({AMR}) Speech Codec; Transcoding Functions}},
  institution = {3rd Generation Partnership Project (3GPP)},
  number      = {TS 26.071},
  year        = {1999}
}

@inproceedings{buthe25bwedemo,
  author    = {B{\"u}the, Jan and Valin, Jean-Marc},
  title     = {{A Lightweight and Robust Method for Blind Wideband-to-Fullband Extension of Speech}},
  booktitle = {Proc. IEEE Workshop on Applications of Signal Processing to Audio and Acoustics (WASPAA)},
  year      = {2025},
  address   = {Tahoe City, CA, USA}
}

@inproceedings{schmidt23lpcgan,
  author    = {Schmidt, Konstantin and Edler, Bernd and Mahmoud, Ahmed Mustafa and Fuchs, Guillaume},
  title     = {{LPC-GAN for Speech Super-Resolution}},
  booktitle = {Proc. European Signal Processing Conference (EUSIPCO)},
  year      = {2023},
  month     = sep,
  address   = {Helsinki, Finland},
  pages     = {346--350}
}

@inproceedings{mustafa21stylemelgan,
  author    = {Mustafa, Ahmed and Pia, Nicola and Fuchs, Guillaume},
  title     = {{StyleMelGAN: An Efficient High-Fidelity Adversarial Vocoder with Temporal Adaptive Normalization}},
  booktitle = {Proc. IEEE ICASSP},
  pages     = {6034--6038},
  year      = {2021},
  month     = jun,
  address   = {Toronto, ON, Canada},
}

@article{nguyen94pqmf,
  title={{Near-perfect-reconstruction pseudo-QMF banks}},
  author={Nguyen, Truong Q},
  journal={IEEE Transactions on Signal Processing},
  volume={42},
  number={1},
  pages={65--76},
  year={1994},
  publisher={IEEE}
}

@inproceedings{wen24ftcsr,
  author    = {Wen, Liang and Wang, Lizhong and Zheng, Yuxing and Shi, Weijing and Choi, Kwang Pyo},
  title     = {{FT-CSR: Cascaded Frequency-Time Method for Coded Speech Restoration}},
  booktitle = {Proc. IEEE International Conference on Multimedia and Expo (ICME)},
  year      = {2024},
  address = {Niagara Falls, ON, Canada},
  month = jul
}

@inproceedings{gupta25ubgan,
  author    = {Gupta, Kishan and Korse, Srikanth and Brendel, Andreas and Pia, Nicola and Fuchs, Guillaume},
  title     = {{UBGAN: Enhancing Coded Speech With Blind and Guided Bandwidth Extension}},
  booktitle = {Proc. IEEE Workshop on Applications of Signal Processing to Audio and Acoustics (WASPAA)},
  year      = {2025},
  address   = {Tahoe City, CA, USA}
}

@inproceedings{su21hifigan,
  author    = {Su, Jiaqi and Wang, Yunyun and Finkelstein, Adam and Jin, Zeyu},
  title     = {{Bandwidth Extension Is All You Need}},
  booktitle = {Proc. IEEE International Conference on Acoustics, Speech and Signal Processing (ICASSP)},
  year      = {2021},
  month     = jun,
  address   = {Toronto, ON, Canada},
  pages     = {696--700}
}

@inproceedings{oord16wavenet,
  author    = {van den Oord, A{\"a}ron and Dieleman, Sander and Zen, Heiga and Simonyan, Karen and Vinyals, Oriol and Graves, Alex and Kalchbrenner, Nal and Senior, Andrew and Kavukçuoğlu, Koray},
  title     = {{WaveNet: A Generative Model for Raw Audio}},
  booktitle = {Proc. ISCA Speech Synthesis Workshop (SSW 9)},
  year      = {2016},
  month     = sep,
  address   = {Sunnyvale, CA, USA},
  pages     = {125}
}

@article{kolbaek16intelligibility,
  author  = {Kolb{\ae}k, Morten and Tan, Zheng-Hua and Jensen, Jesper},
  title   = {{Speech Intelligibility Potential of General and Specialized Deep Neural Network Based Speech Enhancement Systems}},
  journal = {IEEE/ACM Trans. on Audio, Speech, and Language Processing},
  volume  = {25},
  number  = {1},
  pages   = {153--167},
  year    = {2017},
  month   = jan
}

@inproceedings{sivaraman21zeroshot,
  author    = {Sivaraman, Aswin and Kim, Minje},
  title     = {{Zero-Shot Personalized Speech Enhancement Through Speaker-Informed Model Selection}},
  booktitle = {Proc. IEEE Workshop on Applications of Signal Processing to Audio and Acoustics (WASPAA)},
  year      = {2021},
  month     = oct,
  address   = {New Paltz, NY, USA},
  pages     = {171--175}
}

@inproceedings{eskimez22personalizedse,
  author    = {Eskimez, {\c{S}}efik Emre and Yoshioka, Takuya and Wang, Huaming and Wang, Xiaofei and Chen, Zhuo and Huang, Xuedong},
  title     = {{Personalized Speech Enhancement: New Models and Comprehensive Evaluation}},
  booktitle = {Proc. IEEE International Conference on Acoustics, Speech and Signal Processing (ICASSP)},
  year      = {2022},
  month     = may,
  address   = {Singapore},
  pages     = {356--360}
}

@inproceedings{sivaraman20sparsemixture,
  author    = {Sivaraman, Aswin and Kim, Minje},
  title     = {{Sparse Mixture of Local Experts for Efficient Speech Enhancement}},
  booktitle = {Proc. Interspeech},
  year      = {2020},
  month     = oct,
  address   = {Shanghai, China},
  pages     = {4526--4530},
  issn      = {2958-1796}
}

@inproceedings{xu25boneconduction,
  author    = {Xu, Pan and Zhang, Zhongyu and Fu, Zhonghua},
  title     = {{Personalized Bone-Conduction Bandwidth Extension With Speaker Characteristics}},
  booktitle = {Proc. Asia Pacific Signal and Information Processing Association Annual Summit and Conference (APSIPA ASC)},
  year      = {2025},
  pages     = {1128--1133},
  month     = oct,
  address   = {Singapore}  
}

@inproceedings{jang24personalizedcodec,
  author    = {Jang, Inseon and Yang, Haici and Lim, Wootaek and Beack, Seungkwon and Kim, Minje},
  title     = {{Personalized Neural Speech Codec}},
  booktitle = {Proc. IEEE International Conference on Acoustics, Speech and Signal Processing (ICASSP)},
  year      = {2024},
  month     = apr,
  address   = {Seoul, South Korea},
  pages     = {991--995}
}

@inproceedings{delcroix18speakerbeam,
  author    = {Delcroix, Marc and Žmolíková, Kateřina and Kinoshita, Keisuke and Ogawa, Atsunori and Nakatani, Tomohiro},
  title     = {{Single-Channel Target Speaker Extraction and Recognition With SpeakerBeam}},
  booktitle = {Proc. IEEE International Conference on Acoustics, Speech and Signal Processing (ICASSP)},
  year      = {2018},
  month     = apr,
  address   = {Calgary, AB, Canada},
  pages     = {5554--5558}
}

@inproceedings{denisov19endtoend,
  author    = {Denisov, Pavel and Vu, Thang},
  title     = {{End-to-End Multi-Speaker Speech Recognition Using Speaker Embeddings and Transfer Learning}},
  booktitle = {Proc. Interspeech},
  year      = {2019},
  month     = sep,
  address   = {Graz, Austria},
  pages     = {4425--4429},
  doi       = {10.21437/Interspeech.2019-1130}
}

@inproceedings{ding20personalvad,
  author    ={Ding, Shaojin and Wang, Quan and Chang, Shuo-yiin and Wan, Li and Moreno, Ignacio Lopez},
  title     = {{Personal {VAD}: Speaker-Conditioned Voice Activity Detection}},
  booktitle = {Proc. Odyssey 2020: The Speaker and Language Recognition Workshop},
  year      = {2020},
  month     = nov,
  address   = {Tokyo, Japan},
  pages     = {433--439}
}

@inproceedings{medennikov20tsvad,
  author={Medennikov, Ivan and Korenevsky, Maxim and Prisyach, Tatiana and Khokhlov, Yuri and Korenevskaya, Mariya and Sorokin, Ivan and Timofeeva, Tatiana and Mitrofanov, Anton and Andrusenko, Andrei and Podluzhny, Ivan and others},
  title     = {{Target-Speaker Voice Activity Detection: A Novel Approach for Multi-Speaker Diarization in a Dinner Party Scenario}},
  booktitle = {Proc. Interspeech},
  year      = {2020},
  month     = oct,
  address   = {Shanghai, China},
  pages     = {274--278}
}

@inproceedings{cooper20zeroshottts,
  author={Cooper, Erica and Lai, Cheng-I and Yasuda, Yusuke and Fang, Fuming and Wang, Xin and Chen, Nanxin and Yamagishi, Junichi},
  title     = {{Zero-Shot Multi-Speaker Text-to-Speech With State-of-the-Art Neural Speaker Embeddings}},
  booktitle = {Proc. IEEE International Conference on Acoustics, Speech and Signal Processing (ICASSP)},
  year      = {2020},
  month     = may,
  address   = {Barcelona, Spain},
  pages     = {6184--6188}
}

@article{wang24usat,
  author={Wang, Wenbin and Song, Yang and Jha, Sanjay},
  title   = {{USAT: A Universal Speaker-Adaptive Text-to-Speech Approach}},
  journal = {IEEE/ACM Trans. on Audio, Speech, and Language Processing},
  volume  = {32},
  pages   = {2590--2604},
  year    = {2024}
}

@inproceedings{nagrani17voxceleb,
  title={{VoxCeleb: A Large-Scale Speaker Identification Dataset}},
  author={Nagrani, Arsha and Chung, Joon Son and Zisserman, Andrew},
  booktitle = {Proc. Interspeech},
  pages={2616--2620},
  year={2017},
  month     = aug,
  address   = {Stockholm, Sweden},
}

@inproceedings{chung18b_voxceleb2,
  title     = {{VoxCeleb2: Deep Speaker Recognition}},
  author    = {Joon Son Chung and Arsha Nagrani and Andrew Zisserman},
  year      = {2018},
  booktitle = {{Proc. Interspeech 2018}},
  pages     = {1086--1090},
  month     = sep,
  address   = {Hyderabad, India},
}

@article{zmolikova23ntseoverview,
  author={Žmolíková, Kateřina and Delcroix, Marc and Ochiai, Tsubasa and Kinoshita, Keisuke and {\v{C}}ernock{\'y}, Jan and Yu, Dong},
  title   = {{Neural Target Speech Extraction: An Overview}},
  journal = {IEEE Signal Process. Mag.},
  volume  = {40},
  number  = {3},
  pages   = {8--29},
  year    = {2023},
  month   = may
}

@inproceedings{desplanques20ecapatdnn,
  author={Desplanques, Brecht and Thienpondt, Jenthe and Demuynck, Kris},
  title     = {{ECAPA-TDNN: Emphasized Channel Attention, Propagation and Aggregation in {TDNN}-Based Speaker Verification}},
  booktitle = {Proc. Interspeech},
  year      = {2020},
  month     = oct,
  address   = {Shanghai, China},
  pages     = {3830--3834}
}

@techreport{pesq,
  author      = "{ITU-T}",
  title       = "{Wideband Extension to Recommendation P.862 for the Assessment of Wideband Telephone Networks and Speech Codecs}",
  institution = "{International Telecommunication Union}",
  type        = "{ITU-T Rec.}",
  number      = "P.862.2",
  address     = "Geneva, Switzerland",
  year        = "2005",
}

@inproceedings{chinen20visqol,
  author={Chinen, Michael and Lim, Felicia S. C. and Skoglund, Jan and Gureev, Nikita and O'Gorman, Feargus and Hines, Andrew},
  title={{ViSQOL v3: An Open Source Production Ready Objective Speech and Audio Metric}}, 
  booktitle={Proc. Quality of Multimedia Experience (QoMEX)}, 
  year={2020},
  month  = oct,
  address = {Athlone, Ireland}
}

@misc{yamagishi2019vctk,
  author       = {Yamagishi, Junichi and Veaux, Christophe and MacDonald, Kirsten},
  title        = {{CSTR {VCTK} Corpus: English Multi-Speaker Corpus for {CSTR} Voice Cloning Toolkit (Version 0.92)}},
  howpublished = {Centre for Speech Technology Research (CSTR), University of Edinburgh},
  year         = {2019}
}

\end{document}